\documentclass[conference]{IEEEtran}
\usepackage{url}
\usepackage[utf8]{inputenc}
\usepackage{xcolor}
\usepackage{amsmath}

\usepackage{multirow}
\usepackage{amssymb}
\usepackage{cite}
\usepackage{enumitem}
\usepackage{bm}
\usepackage{graphicx}
\usepackage{amsthm}
\usepackage{physics}

\usepackage{comment}

\usepackage[utf8]{inputenc}
\usepackage{pgfplots}
\DeclareUnicodeCharacter{2212}{−}
\usepgfplotslibrary{groupplots,dateplot}
\usetikzlibrary{patterns,shapes.arrows}
\pgfplotsset{compat=newest}

\usepackage{gensymb}

\usepackage[acronyms,nonumberlist,nopostdot,nomain,nogroupskip]{glossaries}
\usepackage{tablefootnote}
\usepackage{booktabs}
\usepackage{tabularx}
\usepackage{epsfig}
\usepackage{tikz}
\usetikzlibrary{external}
\usepackage{pgfplots}
\pgfplotsset{compat=newest} 
\pgfplotsset{plot coordinates/math parser=false}

\usetikzlibrary{plotmarks,patterns,patterns.meta,decorations.pathreplacing,backgrounds,calc,arrows,arrows.meta,spy,matrix}
\usepgfplotslibrary{patchplots,groupplots}
\usepackage{tikzscale}
\usepackage{siunitx}
\usepackage{tablefootnote}

\usepackage{multirow}
\usepackage{algorithm2e, setspace}
\SetAlCapNameFnt{\footnotesize}
\SetAlCapFnt{\footnotesize}
\RestyleAlgo{ruled}
\SetKwComment{Comment}{/* }{ */}
\usepackage{algpseudocode}

\usepackage[font=scriptsize]{subcaption}
\usepackage[font=footnotesize]{caption}

\usepackage{mathtools}

\usepackage{dblfloatfix}    
\usepackage{colortbl}
\usepackage{flushend}
\usepackage{footnote}

\usepackage{lipsum,afterpage,refcount}

\newacronym{3gpp}{3GPP}{3rd Generation Partnership Project}
\newacronym{adc}{ADC}{Analog to Digital Converter}
\newacronym{5g}{5G}{5th generation}
\newacronym{6g}{6G}{6th generation}
\newacronym{ai}{AI}{Artificial Intelligence}
\newacronym{aimd}{AIMD}{Additive Increase Multiplicative Decrease}
\newacronym{am}{AM}{Acknowledged Mode}
\newacronym{amc}{AMC}{Adaptive Modulation and Coding}
\newacronym{aqm}{AQM}{Active Queue Management}
\newacronym{awgn}{AGWN}{Additive White Gaussian Noise}
\newacronym{balia}{BALIA}{Balanced Link Adaptation}
\newacronym{bdp}{BDP}{Bandwidth-Delay Product}
\newacronym{bf}{BF}{beamforming}
\newacronym{cc}{CC}{Congestion Control}
\newacronym{cdf}{CDF}{Cumulative Distribution Function}
\newacronym{cn}{CN}{Core Network}
\newacronym{cqi}{CQI}{Channel Quality Information}
\newacronym{cp}{CP}{Control Plane}
\newacronym{csirs}{CSI-RS}{Channel State Information - Reference Signal}
\newacronym{dc}{DC}{Dual Connectivity}
\newacronym{rb}{RB}{Resource Block}
\newacronym{dce}{DCE}{Direct Code Execution}
\newacronym{d2ss}{D2SS}{Direct to Satellite Service}
\newacronym{dci}{DCI}{Downlink Control Information}
\newacronym{udp}{UDP}{User Datagram Protocol}
\newacronym{dl}{DL}{downlink}
\newacronym{fcfs}{FCFS}{first-come-first-served}
\newacronym{dmr}{DMR}{Deadline Miss Ratio}
\newacronym{fspl}{FSPL}{free-space path loss}
\newacronym{dmrs}{DMRS}{DeModulation Reference Signal}
\newacronym{e2e}{E2E}{End-to-End}
\newacronym{ppp}{PPP}{Poission Point Process}
\newacronym{aoi}{AoI}{Area of Interest}
\newacronym{cpu}{CPU}{Central Processing Unit}
 \newacronym{gpu}{GPU}{Graphics Processing Unit}
 \newacronym{tpu}{TPU}{Tensor Processing Unit}
\newacronym{si}{SI}{Study Item}
\newacronym{ecn}{ECN}{Explicit Congestion Notification}
\newacronym{edf}{EDF}{Earliest Deadline First}
\newacronym{enb}{eNB}{eNodeB}
\newacronym{epc}{EPC}{Evolved Packet Core}
\newacronym{es}{ES}{Edge Server}
\newacronym{cav}{CAV}{Connected and Autonomous Vehicle}
\newacronym{fdma}{FDMA}{Frequency Division Multiple Access}
\newacronym{fdd}{FDD}{Frequency Division Duplexing}
\newacronym{upa}{UPA}{Uniform Planar Array}
\newacronym{car}{CAR}{Circular Aperture Reflector }
\newacronym[firstplural=Radio Access Technologies (RATs)]{rat}{RAT}{Radio Access Technology}
\newacronym[firstplural=Radio Access Technology (RTs)]{rt}{RT}{Radio Technology}
\newacronym{fs}{FS}{Fast Switching}
\newacronym{isd}{ISD}{inter-site distance}
\newacronym{ftp}{FTP}{File Transfer Protocol}
\newacronym{gnb}{gNB}{Next Generation Node Base}
\newacronym{harq}{HARQ}{Hybrid Automatic Repeat reQuest}
\newacronym{hetnet}{HetNet}{Heterogeneous Network}
\newacronym{hh}{HH}{Hard Handover}
\newacronym{hol}{HOL}{Head-of-Line}
\newacronym{ia}{IA}{Initial Access}
\newacronym{imt}{IMT}{International Mobile Telecommunication}
\newacronym{iot}{IoT}{Internet of Things}
\newacronym{los}{LOS}{Line of Sight}
\newacronym{lte}{LTE}{Long Term Evolution}
\newacronym{m2m}{M2M}{Machine to Machine}
\newacronym{mac}{MAC}{Medium Access Control}
\newacronym{mc}{MC}{Multi-Connectivity}
\newacronym{mcs}{MCS}{Modulation and Coding Scheme}
\newacronym{mec}{MEC}{Mobile Edge Cloud}
\newacronym{mi}{MI}{Mutual Information}
\newacronym{mimo}{MIMO}{Multiple Input Multiple Output}
\newacronym{mmwave}{mmWave}{millimeter wave}
\newacronym{mptcp}{MPTCP}{Multipath TCP}
\newacronym{mr}{MR}{Maximum Rate}
\newacronym{mss}{MSS}{Maximum Segment Size}
\newacronym{mtd}{MTD}{Machine-Type Device}
\newacronym{mtu}{MTU}{Maximum Transmission Unit}
\newacronym{nfv}{NFV}{Network Function Virtualization}
\newacronym{vnf}{VNF}{Virtualization Network Function}
\newacronym{gv}{GV}{ground vehicle}
\newacronym{gvs}{GVs}{ground vehicles}
\newacronym{vec}{VEC}{Vehicular Edge Computing}
\newacronym{sdn}{SDN}{Software Defined Networking}
\newacronym{nlos}{NLOS}{Non Line of Sight}
\newacronym{nlosb}{NLOSb}{Building Non Line of Sight}
\newacronym{nlosv}{NLOSv}{Vehicle Non Line of Sight}
\newacronym{nr}{NR}{New Radio}
\newacronym{ofdm}{OFDM}{Orthogonal Frequency Division Multiplexing}
\newacronym{pdcch}{PDCCH}{Physical Downlink Control Channel}
\newacronym{pdcp}{PDCP}{Packet Data Convergence Protocol}
\newacronym{pdsch}{PDSCH}{Physical Downlink Shared Channel}
\newacronym{pdu}{PDU}{Packet Data Unit}
\newacronym{pf}{PF}{Proportional Fair}
\newacronym{pgw}{PGW}{Packet Gateway}
\newacronym{phy}{PHY}{Physical}
\newacronym{pbch}{PBCH}{Physical Broadcast Channel}
\newacronym[plural=\gls{mme}s,firstplural=Mobility Management Entities (MMEs)]{mme}{MME}{Mobility Management Entity}
\newacronym{prb}{PRB}{Physical Resource Block}
\newacronym{pss}{PSS}{Primary Synchronization Signal}
\newacronym{pucch}{PUCCH}{Physical Uplink Control Channel}
\newacronym{pusch}{PUSCH}{Physical Uplink Shared Channel}
\newacronym{rach}{RACH}{Random Access Channel}
\newacronym{ran}{RAN}{Radio Access Network}
\newacronym{red}{RED}{Random Early Detection}
\newacronym{rf}{RF}{Radio Frequency}
\newacronym{rlc}{RLC}{Radio Link Control}
\newacronym{rlf}{RLF}{Radio Link Failure}
\newacronym{rrc}{RRC}{Radio Resource Control}
\newacronym{rrm}{RRM}{Radio Resource Management}
\newacronym{rr}{RR}{Round Robin}
\newacronym{rs}{RS}{Remote Server}
\newacronym{rsrp}{RSRP}{Reference Signal Received Power}
\newacronym{rss}{RSS}{Received Signal Strength}
\newacronym{rtt}{RTT}{Round Trip Time}
\newacronym{rw}{RW}{Receive Window}
\newacronym{rx}{RX}{Receiver}
\newacronym{sa}{SA}{standalone}
\newacronym{sack}{SACK}{Selective Acknowledgment}
\newacronym{sap}{SAP}{Service Access Point}
\newacronym{sch}{SCH}{Secondary Cell Handover}
\newacronym{scoot}{SCOOT}{Split Cycle Offset Optimization Technique}
\newacronym{sdma}{SDMA}{Spatial Division Multiple Access}
\newacronym{sinr}{SINR}{Signal to Interference plus Noise Ratio}
\newacronym{sm}{SM}{Saturation Mode}
\newacronym{snr}{SNR}{Signal-to-Noise Ratio}
\newacronym{son}{SON}{Self-Organizing Network}
\newacronym{ss}{SS}{Synchronization Signal}
\newacronym{srs}{SRS}{Sounding Reference Signal}
\newacronym{sss}{SSS}{Secondary Synchronization Signal}
\newacronym{tb}{TB}{Transport Block}
\newacronym{tcp}{TCP}{Transmission Control Protocol}
\newacronym{tdd}{TDD}{Time Division Duplexing}
\newacronym{tdma}{TDMA}{Time Division Multiple Access}
\newacronym{tfl}{TfL}{Transport for London}
\newacronym{tm}{TM}{Transparent Mode}
\newacronym{prr}{PRR}{Packet Reception Ratio}
\newacronym{trp}{TRP}{Transmitter Receiver Pair}
\newacronym{tti}{TTI}{Transmission Time Interval}
\newacronym{ttt}{TTT}{Time-to-Trigger}
\newacronym{tx}{TX}{Transmitter}
\newacronym{ue}{UE}{User Equipment}
\newacronym{ul}{UL}{uplink}
\newacronym{uml}{UML}{Unified Modeling Language}
\newacronym{um}{UM}{Unacknowledged Mode}
\newacronym{utc}{UTC}{Urban Traffic Control}
\newacronym{vm}{VM}{Virtual Machine}
\newacronym{rsrq}{RSRQ}{Reference Signal Received Quality}
\newacronym{rssi}{RSSI}{Received Signal Strength Indicator}
\newacronym{crs}{CRS}{Cell Reference Signal}
\newacronym{v2v}{V2V}{Vehicle-to-Vehicle}
\newacronym{v2i}{V2I}{Vehicle-to-Infrastructure}
\newacronym{v2n}{V2N}{Vehicle-to-Network}
\newacronym{v2x}{V2X}{Vehicle-to-Everything}
\newacronym{vn}{VN}{Vehicular Node}
\newacronym{dsrc}{DSRC}{Dedicated Short Range Communication}
\newacronym{ci}{CI}{context information}
\newacronym{voi}{VoI}{value of information}
\newacronym{gps}{GPS}{Global Positioning System}
\newacronym{qos}{QoS}{Quality of Service}
\newacronym{qoe}{QoE}{Quality of Experience}
\newacronym{ml}{ML}{Machine Learning}
\newacronym{ahp}{AHP}{Analytic Hierarchy Process}
\newacronym{lidar}{LiDAR}{Light Detection and Ranging}
\newacronym{sumo}{SUMO}{Simulation of Urban MObility}
\newacronym{wave}{WAVE}{Wireless Access in Vehicular Environment}
\newacronym{c-its}{C-ITS}{Connected Intelligent Transportation System}
\newacronym{dash}{DASH}{Dynamic Adaptive Streaming over HTTP}
\newacronym{http}{HTTP}{HyperText Transfer Protocol}
\newacronym{nt}{NT}{Non-Terrestrial}
\newacronym{ntc}{NTC}{non-terrestrial communication}
\newacronym{ntn}{NTN}{Non-Terrestrial Network}
\newacronym{tn}{TN}{Terrestrial Network}
\newacronym{ta}{TA}{Timing Advance}
\newacronym{hap}{HAP}{High Altitude Platform}
\newacronym{leo}{LEO}{Low Earth Orbit}
\newacronym{meo}{MEO}{Medium Earth Orbit}
\newacronym{geo}{GEO}{Geostationary Earth Orbit}
\newacronym{uav}{UAV}{Unmanned Aerial Vehicle}
\newacronym{nsat}{nSAT}{Nanosatellite}
\newacronym{ehf}{EHF}{extremely high-frequency}
\newacronym{ioe}{IoE}{Internet of Everyone}
\newacronym{gan}{GaN}{Gallium Nitride}
\newacronym{tle}{TLE}{two-line element}
\newacronym{ecdf}{ECDF}{Empirical Cumulative Distribution Function}
\newacronym{fifo}{FIFO}{First-Input First-Output}
\newacronym{gnss}{GNSS}{Global Navigation Satellite System}
\newacronym{essa}{ESSA}{Enhanced Synchronized Slot Allocation}
\newacronym{pdcc}{PDCC}{Physical Downlink Control Channel}
\newacronym{milp}{MILP}{Mixed Integer Linear Programming}
\newacronym{iqr}{IQR}{Interquartile Range}
\newacronym{gp}{GP}{Guard Period}
\newacronym{ncs}{NCS}{Networked Control System}
\newacronym{iiot}{IIoT}{Industrial Internet of Things}
\newacronym{jcc}{JCC}{Joint Communication and Control}
\newacronym{agv}{AGV}{Automated Guided Vehicle}
\newacronym{mse}{MSE}{Mean Squared Error}
\newacronym{ros2}{ROS 2}{Robot Operating System 2}

\usepackage{tikz}
\usepackage{pgfplots}
\usepackage{tikzscale}
\usepackage{tikz-qtree}

\pgfplotsset{compat=newest}
\pgfplotsset{plot coordinates/math parser=false}
\pgfplotsset{every axis/.append style={
                    label style={font=\scriptsize},
                    tick label style={font=\scriptsize},
                    legend style={font=\scriptsize}
                    }}
\usetikzlibrary{plotmarks,shapes,patterns,decorations.pathreplacing,backgrounds,calc,arrows,arrows.meta,spy,matrix,shadows,trees,positioning,fit}
\usepgfplotslibrary{patchplots,groupplots}

\tikzstyle{startstop} = [rectangle, rounded corners, minimum width=2cm, minimum height=0.5cm,text centered, draw=black]
\tikzstyle{io} = [trapezium, trapezium left angle=70, trapezium right angle=110, minimum width=3cm, minimum height=1cm, text centered, draw=black]
\tikzstyle{process} = [rectangle, minimum width=2cm, minimum height=0.5cm, text centered, draw=black, alignb=center]
\tikzstyle{decision} = [ellipse, minimum width=2cm, minimum height=1cm, text centered, draw=black]
\tikzstyle{arrow} = [thick,<->,>=stealth]
\tikzstyle{line} = [thick,>=stealth]
\tikzstyle{darrow} = [thick,<->,>=stealth,dashed]
\tikzstyle{sarrow} = [thick,->,>=stealth]
\tikzstyle{larrow} = [line width=0.1mm,dashdotted,->,>=stealth]

\pgfkeys{/pgf/number format/.cd,1000 sep={\,}} 

\makeatletter
\def\grd@save@target#1{%
  \def\grd@target{#1}}
\def\grd@save@start#1{%
  \def\grd@start{#1}}
\tikzset{
  grid with coordinates/.style={
    to path={%
      \pgfextra{%
        \edef\grd@@target{(\tikztotarget)}%
        \tikz@scan@one@point\grd@save@target\grd@@target\relax
        \edef\grd@@start{(\tikztostart)}%
        \tikz@scan@one@point\grd@save@start\grd@@start\relax
        \draw[minor help lines] (\tikztostart) grid (\tikztotarget);
        \draw[major help lines] (\tikztostart) grid (\tikztotarget);
        \grd@start
        \pgfmathsetmacro{\grd@xa}{\the\pgf@x/1cm}
        \pgfmathsetmacro{\grd@ya}{\the\pgf@y/1cm}
        \grd@target
        \pgfmathsetmacro{\grd@xb}{\the\pgf@x/1cm}
        \pgfmathsetmacro{\grd@yb}{\the\pgf@y/1cm}
        \pgfmathsetmacro{\grd@xc}{\grd@xa + \pgfkeysvalueof{/tikz/grid with coordinates/major step x}}
        \pgfmathsetmacro{\grd@yc}{\grd@ya + \pgfkeysvalueof{/tikz/grid with coordinates/major step y}}
        \foreach \x in {\grd@xa,\grd@xc,...,\grd@xb}
        \node[anchor=north] at (\x,\grd@ya) {\pgfmathprintnumber{\x}};
        \foreach \y in {\grd@ya,\grd@yc,...,\grd@yb}
        \node[anchor=east] at (\grd@xa,\y) {\pgfmathprintnumber{\y}};
      }
    }
  },
  minor help lines/.style={
    help lines,
    gray,
    line cap =round,
    xstep=\pgfkeysvalueof{/tikz/grid with coordinates/minor step x},
    ystep=\pgfkeysvalueof{/tikz/grid with coordinates/minor step y}
  },
  major help lines/.style={
    help lines,
    line cap =round,
    line width=\pgfkeysvalueof{/tikz/grid with coordinates/major line width},
    xstep=\pgfkeysvalueof{/tikz/grid with coordinates/major step x},
    ystep=\pgfkeysvalueof{/tikz/grid with coordinates/major step y}
  },
  grid with coordinates/.cd,
  minor step x/.initial=.5,
  minor step y/.initial=.2,
  major step x/.initial=1,
  major step y/.initial=1,
  major line width/.initial=1pt,
}
\makeatother

\newlength\fheight
\newlength\fwidth

\definecolor{steelblue}{RGB}{176,196,222}

\usepackage[capitalize]{cleveref}
\crefname{section}{Sec.}{Secs.}
\usetikzlibrary{decorations}

\usepackage[utf8]{inputenc}
\usepackage{pgfplots}
\DeclareUnicodeCharacter{2212}{−}
\usepgfplotslibrary{groupplots,dateplot}
\usetikzlibrary{patterns,shapes.arrows}
\pgfplotsset{compat=newest}

\begin{document}
\bstctlcite{IEEEexample:BSTcontrol}

\title{Network-Aware Control of \glspl{agv} in an Industrial Scenario: A Simulation Study Based on \\ROS 2 and Gazebo\vspace{-0.1cm}}
\author{\IEEEauthorblockN{\textit{(Invited Paper)\vspace{0.3cm}} \\ Filippo Bragato$^{\star }$, Tullia Fontana$^{\star }$, Marco Giordani$^{\star }$, Malte Schellmann$^{\dagger }$, Josef Eichinger$^{\dagger }$, Michele Zorzi$^{\star }$}
\IEEEauthorblockA{$^{\star }$ WiLab/CNIT and University of Padova, Italy.
Email:	{\{bragato,fontana,giordani,zorzi\}@dei.unipd.it}\\
$^{\dagger }$ Huawei Technologies, Munich Research Center, Germany. \{malte.schellmann,joseph.eichinger\}@huawei.com.}\vspace{-1cm}}

\maketitle

\glsresetall

\begin{abstract}
\gls{ncs} is a paradigm where sensors, controllers, and actuators communicate over a shared network. One promising application of NCS is the control of \glspl{agv} in the industrial environment, for example to transport goods efficiently and to autonomously follow predefined paths or routes. 
In this context, communication and control are tightly correlated, a paradigm referred to as \gls{jcc}, since network issues such as delays or errors can lead to significant deviations of the AGVs from the planned trajectory.
In this paper, we present a simulation framework based on Gazebo and \gls{ros2} to simulate and visualize, respectively, the complex interaction between the control of AGVs and the underlying communication network.
This framework explicitly incorporates communication metrics, such as delay and packet loss, and control metrics, especially the \gls{mse} between the optimal/desired and actual path of the AGV in response to driving commands.
Our results shed light into the correlation between the network performance, particularly \gls{prr}, and accuracy of control.
\end{abstract}

\glsresetall

\begin{IEEEkeywords}
\Glspl{agv}; \gls{ncs}; \gls{jcc}; \gls{ros2}; Gazebo. 
\end{IEEEkeywords}

\begin{tikzpicture}[remember picture,overlay]
  \node[anchor=north,yshift=-10pt] at (current page.north) {\parbox{\dimexpr\textwidth-\fboxsep-\fboxrule\relax}{
      \centering\footnotesize This paper has been accepted for publication at IEEE International Symposium on Personal, \\Indoor and Mobile Radio Communications (PIMRC). 2025 ©IEEE.\\
      Please cite it as: F. Bragato, T. Fontana, M. Giordani, M. Schellmann, J. Eichinger, M. Zorzi, “Network-Aware Control of AGVs in an Industrial Scenario: A Simulation Study Based on ROS 2 and Gazebo,” IEEE International Symposium on Personal, Indoor and Mobile Radio Communications (PIMRC), Istanbul, Türkiye, 2025.\\
      }};
\end{tikzpicture}

\glsresetall

\section{Introduction}
\label{sec:intro}
\gls{ncs} for inter-machine \gls{iiot} is a paradigm that involves using a wireless communication network to connect sensors, actuators, and controllers within a factory environment \cite{NCS_survey_2024}. 
Specifically, the goals of \gls{ncs} are to reduce deployment costs, improve system performance, and achieve greater flexibility in industrial settings~\cite{wireless_vs_wired_2009}. 
Notably, in the paradigm of Industry 4.0~\cite{2019_industry_4.0}, one promising application of NCS 
is the remote control of teleoperated \glspl{agv}, to navigate and perform tasks in dynamic environments such as warehouses and production lines~\cite{2022_case_study_agv}.

In practice, supporting real-time control over a wireless communication network is challenging~\cite{cavallero2023new}. Specifically, unlike wired systems, wireless networks are subject to several impairments such as non-negligible transmission delays and errors that may degrade the performance of the control system \cite{Zhang2017Analysis,Hespanha2007A}. 
In early NCSs, communication and control modules were implemented separately and independently of the actual users' applications. In other words, the communication system and the machine control system were basically considered as ``black boxes'' for each other. A common strategy was to analyze communication conditions in the worst-case scenario, thereby setting a lower bound on the control performance. 
More recently, the research community has started to explore a new approach to integrate communication and control in NCS, a paradigm referred to as \gls{jcc}: if communication is not properly addressed in the design of \gls{ncs}, control of machines may be compromised~\cite{survey_NCS}. 
For this co-design problem, 
the most critical variables are the packet delay, the \gls{prr}~\cite{NCS_survey_2024}, and the underlying control error~\cite{Wang2020Stability,Tan2023Integrated,cuozzo2022enabling}. For example, for the remote control of an AGV, communication delays may cause the AGV to respond too late to commands, resulting in navigation errors or, in the worst case, accidents.
As such, the communication network and the control system are to be jointly optimized to address certain~requirements.

Different simulators exist to optimize and validate \gls{ncs} solutions prior to full-scale deployment~\cite{López2021A}.
For example, Gazebo is a dynamic simulator with the ability to visualize industrial scenarios in 3D with a high degree of fidelity. \gls{ros2} is another important tool for building and simulating industrial machine applications in 2D.
In this sense, integrating Gazebo and \mbox{\gls{ros2}} together permits simulation and visualization, respectively, of the complex interactions between the communication network and the industrial control system.
For example, Yumbla \emph{et al.}, in~\cite{Yumbla2025An}, used \gls{ros2} and Gazebo to create a framework to coordinate robotic agents in a collaborative environment. Similarly, Mengacci \emph{et al.}, in~\cite{Mengacci2021An}, developed a ROS-Gazebo toolbox to design planning and control strategies for mobile robots in a simulated and scalable environment before direct testing of the real robot applications.
However, prior work does not generally take into account the effects of the wireless channel on the control system.

\begin{figure*}[t!]
    \centering
    \includegraphics[width=0.9\linewidth]{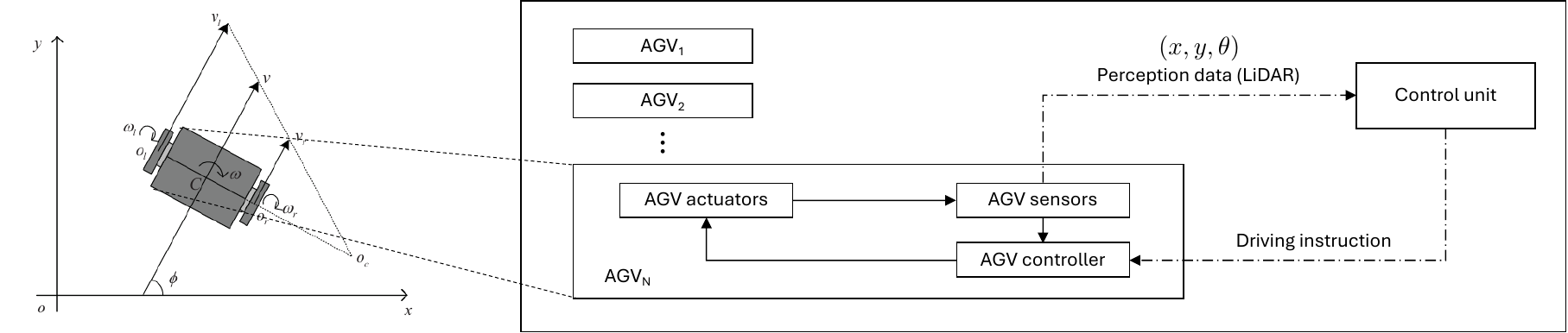}
    \caption{Diagram of the two-wheel differential-drive AGV model (left) and the NCS case study for the remote control of AGVs (right).\vspace{-0.5cm}}
    \label{fig:AGV_scenario}
\end{figure*}

Starting from these contributions, in this paper we propose a new simulation framework that integrates \gls{ros2} and Gazebo to evaluate and improve the reliability and accuracy of an \gls{ncs} industrial control system.
Specifically, we focus on the control of AGVs moving along a factory floor following a predefined path based on driving commands they receive from a remote server via wireless communication. To this end, Gazebo is used to simulate the physics of the AGVs, while \gls{ros2} is used to control the speed and trajectory of the AGVs through interaction with the network.
The factory layout is characterized by a two-state Markov wireless channel, modeling the transitions between good and bad network conditions. 
We validate our simulation framework by measuring the impact of network metrics, including the communication delay and the \gls{prr}, on the controller performance, measured in terms of the \gls{mse} between the optimal/desired and the actual path of the AGV in response to driving commands. 
We demonstrate that the stability of the control system depends on the intricate relationship between the controller and the network, particularly the \gls{prr}.
Moreover, we highlight that there are some conditions where the controller is more sensitive to network problems, and show how its parameters can be adjusted to optimize the system effectively.



The paper is organized as follows. 
Sec.~\ref{sec:model} presents the system model, including both network and control parameters. Sec.~\ref{sec:simulation} describes our simulation framework and methodology. Sec.~\ref{sec:perf-eval} presents our simulation results. Finally, Sec.~\ref{sec:conclusions} concludes this study with future research directions.

\section{System Model}
\label{sec:model}

In this section, we describe our AGV (Sec.~\ref{sub:scenario}), control (Sec.~\ref{sub:control}), and channel (Sec.~\ref{sub:channel-model}) models.

\subsection{Industrial Scenario and AGV Model}
\label{sub:scenario}
We consider a smart factory where AGVs transport raw materials and finished products to/from a warehouse from/to the factory floor, respectively.
\Glspl{agv} integrate advanced sensing, control, and communication technologies to operate efficiently~\cite{AGV_components_2008,2022_case_study_agv}:
Specifically, they are equipped with sensors, such as \gls{lidar} sensors,
for localization and perception of the environment, and to drive along predefined paths. 
Our working assumption is that AGVs can drive autonomously via an onboard control system on very simple tracks (e.g., following a line on the floor). However, in case they lose the track, or face unforeseen events (e.g., blockages or obstacles on the track), \glspl{agv} can be controlled remotely via driving commands received from an external control unit. This is similar to the link scheduling problem for robotic motion planning presented in~\cite{one6gwhite}.
\glspl{agv} use 5G wireless networks to interact with the control unit, which may be prone to communication delay or errors.
The \gls{agv} control unit handles the AGV dynamics, as described in Sec.~\ref{sub:control}.


In this paper we consider a two-wheel differential-drive AGV model.
The main physical parameters of the \gls{agv} are illustrated in Fig.~\ref{fig:AGV_scenario} (left) and defined below:
\begin{itemize}
    \item {Wheel centers}: $O_l$ and $O_r$ represent the left and right wheel centers, respectively.
    \item {Distance}: $\rho$ is the distance between the wheels.
    \item {Mass center}: $C$ is the \gls{agv}'s mass center.
    \item {Velocities}: $v_l$ and $v_r$ are the linear velocities of the left and right wheels, respectively. Therefore, the \gls{agv}'s linear ($v$) and angular ($\omega$) velocities can be written as
    \begin{equation}
        v = \left({v_l + v_r}\right)/{2},
    \end{equation}
    \begin{equation}
         \omega = \left({v_l - v_r}\right)/{\rho}.
    \end{equation}
\end{itemize}

The AGV has a caster wheel at the front and two wheels at the back.
The caster wheel is a passive wheel that rotates freely and is used to stabilize the \gls{agv} when moving. The two wheels at the back are the drive wheels, which are connected to the motors that drive the \gls{agv}. The motors are controlled by a controller, which receives velocity commands from the control system, and sends the appropriate signals to the motors to drive the \gls{agv} accordingly.

This model will be implemented in Gazebo and \gls{ros2},
as described in Sec.~\ref{sec:simulation}.


\subsection{Control Model}
\label{sub:control}

The trajectory of the AGV can be represented by coordinates $(x, y,z)$. 
The \gls{agv} can either rotate or move at a predefined (but possibly variable) speed by changing its location and orientation.

The \gls{agv} controller,  illustrated in Fig.~\ref{fig:AGV_scenario} (right),  estimates the current position of the \gls{agv} using perception data obtained via a \gls{lidar}-based localization system, to understand the surrounding environment and make driving decisions accordingly.
The objective of the controller is to drive the \gls{agv} to its intended destination via an optimal path $p^*$. In our model, the shape of the trajectory is an eight shape, which is a common choice in the control literature.
This shape can be described by the following equations:
\begin{align}
    x^*(t) &= A \sin(t); \\
    y^*(t) &= B \sin(t) \cos(t),
\end{align}
where $A$ and $B$ are the amplitudes of the trajectory in the $x$ and $y$ directions, respectively.

Specifically, the controller sends periodic driving instructions to the \gls{agv}, 
i.e., the linear and angular velocities (represented by vectors in the $\{x,y,z\}$ directions) that the AGV must maintain to follow $p^*$.
When a new instruction is received, the \gls{agv} interrupts the current instruction, and implements the new~one.

The controller implements a PID model.
It consists of three components: the Proportional (P) term, which acts as a gain factor to calculate the error, or deviation, between the target value (i.e., trajectory) and the current value; the Integral (I) term, which accumulates the error over time; and the Derivative (D) term, which measures the rate of change of the error. 
These three components are combined together as a weighted sum to compute the optimal driving instructions that minimize the angular error $\theta_e$ and the linear error $d_e$ between the actual position of the \gls{agv}, given by coordinates $x(\cdot)$ and $y(\cdot)$, and the desired position after an interval $\tau$, given by coordinates $x^*(\cdot)$ and $y^*(\cdot)$, i.e., 
\begin{equation}
    \resizebox{.9\hsize}{!}{$\theta_e(t,\tau) = \text{atan2}\Big(y^*(t+\tau) - y(t), x^*(t+\tau) - x(t)\Big) - \theta(t);$}
\end{equation}
\begin{equation}
    d_e(t,\tau) = \sqrt{\Big(x^*(t+\tau) - x(t)\Big)^2 + \Big(y^*(t+\tau) - y(t)\Big)^2}.
\end{equation}
Actually, for the control of the \gls{agv}, two PID models are used: one for the linear velocity in the $x$ direction, and one for the angular velocity in the $z$ direction.

In order to evaluate the performance of the control system, we define the normalized error function $\varepsilon$, i.e., the \gls{mse} relative to the linear error $d_e$,~as 
\begin{equation} \label{eq:normalized_error}
    \varepsilon = \frac{1}{t_1-t_0} \int_{t_0}^{t_1} \left\| \Big(p^*(t),p(t)\Big) \right\|_2 dt,
\end{equation}
where $||\cdot||_2$ is the Euclidean distance between the optimal/desired path $p^*$ of the AGV and the actual path $p$ of the AGV based on the driving instructions received from the controller.\footnote{Notice that $\varepsilon$ is a function of $d_e$ since $\left\| \big(p^*(t),p(t)\big) \right\|_2 = d_e(t,0)$.} The goal of the control system is to minimize~$\varepsilon$.
This is an example of JCC in the NCS scenario, since communication and control are tightly correlated: communication errors between the AGV and the controller may delay driving instructions, and deteriorate the performance of the~controller.


\subsection{Channel Model}
\label{sub:channel-model}

The 5G channel between the \gls{agv} and the control unit is modeled as a two-state Markov process, to capture realistic and stochastic fluctuations of the channel quality due to environmental dynamics, mobility, and/or interference in the factory floor.
Specifically, the channel alternates between a ``good'' ($G$) and a ``bad'' ($B$)  state depending on the \gls{snr}: the $G$ ($B$) state represents favorable (adverse) propagation conditions, characterized by high (low) \gls{prr} and low (high) communication delay $\delta$.
The state also affects the control system, since missing or delayed data, i.e., driving instructions, from the control unit 
in the $B$ state can affect the ability of the controller to follow the desired~path $p^*$, with negative implications on the resulting control error. In general, the Markov process is represented by stationary probabilities $P[G]$ and $P[B]$.

\section{Simulation Framework}
\label{sec:simulation}
In this section, we present our \gls{ncs} simulation framework to be used to measure the impact of the communication network on the control system, and vice versa. It consists of a Gazebo (Sec.~\ref{sub:gazebo}) and a \gls{ros2} (Sec.~\ref{sub:ros2}) module.
\subsection{Gazebo Module}
\label{sub:gazebo}
The Gazebo \cite{gazebo} module simulates the actual physics of the \gls{agv}.
It consists of a world frame with a ground plane surrounded by four walls forming a square box, and an \gls{agv} model as described in Sec.~\ref{sub:scenario}. 
At the beginning of the simulation, the \gls{agv} is placed at the center of the world frame with an orientation parallel to the x-axis, at coordinates $(x,y,z)=(0, 0, 0)$.


The main component in Gazebo is the engine that computes and simulates the dynamics of the objects, especially the \glspl{agv}, in the world frame. The engine operates at a fixed {update rate} $R$.
To this end, it solves the equations of motion for the objects based on the laws of physics. Specifically, it calculates the position and orientation of the objects at each frame, and updates them according to the driving instructions they receive from the control unit. We use the DART engine, which is the default physics engine in~Gazebo.

\begin{table}[t!]
\vspace{0.2cm}
\scriptsize
\centering
\setlength{\tabcolsep}{2pt}
\renewcommand{\arraystretch}{1.2}
\begin{tabular}{|l|l|c|}
\hline
    \gls{ros2} topic & Gazebo topic & Direction \\\hline
    \verb|/vehicle_blue/cmd_vel| & \verb|/model/vehicle_blue/cmd_vel| & $\rightarrow$\\\hline
    \verb|/vehicle_blue/odometry| & \verb|/model/vehicle_blue/odometry| & $\leftarrow$ \\\hline
    \verb|/clock| & \verb|/clock| & $\leftarrow$ \\\hline
    \verb|/vehicle_blue/lidar| & \verb|/lidar| & $\leftarrow$ \\\hline
\end{tabular}
\caption{The topics used by the \texttt{ros\_gz\_bridge} to interconnect the Gazebo and \gls{ros2} modules.\vspace{-0.5cm}}
\label{tab:bridge_topics}
\end{table}

The control system determines the optimal velocity commands for the \gls{agv} based on \gls{ros2} (see Sec.~\ref{sub:ros2}). To permit the two systems to communicate, we use \texttt{ros\_gz\_bridge}, as reported in Tab. \ref{tab:bridge_topics}. This package provides a set of plugins that can be used to publish and subscribe to topics in Gazebo from \gls{ros2} and vice versa, namely the \texttt{/cmd\_vel} topic (for control commands to the \gls{agv}),  and the \texttt{/lidar} topic (for the position and orientation of the \gls{agv} based on LiDAR measurements).
In other words, the bridge subscribes to a \gls{ros2} (Gazebo) source topic, and publishes messages into a Gazebo (\gls{ros2}) destination topic.
Additionally, the evolution of time in the simulation is controlled and governed by Gazebo. As such, every other component, including the control system, needs to be synchronized with Gazebo. To achieve this synchronization, we exploit the \texttt{/clock} topic of Gazebo, which publishes the current simulation time.

The physics of the \gls{agv} are simulated via Gazebo using the \texttt{ diff\_drive\_plugin}, which needs to be configured with the following parameters:
\begin{itemize}
    \item \texttt{left\_wheel\_joint}: The joint that connects the left wheel to the chassis.
    \item \texttt{right\_wheel\_joint}: The  joint that connects the right wheel to the chassis.
    \item \texttt{wheel\_separation}: The distance between wheels.
    \item \texttt{wheel\_radius}: The radius of the wheels.
    \item \texttt{lidar\_publish\_frequency}: The rate at which \gls{lidar} measurements are published.

\end{itemize}

The \texttt{diff\_drive\_plugin} subscribes to the \texttt{/cmd\_vel} topic, which receives the velocity commands for the \gls{agv}, and publishes the \gls{lidar} measurements, i.e.,  the position and orientation of the \gls{agv} in the world frame, on the \texttt{/lidar} topic.
In this way, we can control the \gls{agv} by publishing velocity commands to the \texttt{/cmd\_vel} topic, while localization is performed through a \gls{lidar}-based system.

\subsection{\gls{ros2} Module}
\label{sub:ros2}
\gls{ros2}~\cite{ros2} provides a set of software libraries and tools for building and simulating robot applications in 2D.
\gls{ros2} operates in a graph-like structure, which consists of nodes, topics, services, parameters, and actions. Specifically, multiple nodes can publish and subscribe to the same topic, and each node subscribed to a topic receives the messages published on the topic by the other nodes. The communication between nodes is asynchronous, meaning that they do not need to wait for each other to communicate, and the communication is decoupled from the computation.

In our setup, Gazebo is responsible for the simulation of the physics of the \gls{agv}, while \gls{ros2} is responsible for the control of the \gls{agv} itself. The control system is implemented as a \gls{ros2} node that subscribes to the \texttt{/lidar} topic to get the position of the \gls{agv}, and publishes the velocity commands to a topic to adjust the \gls{agv} trajectory.

\begin{figure}[t!]
    \centering
    \includegraphics[width=0.8\linewidth]{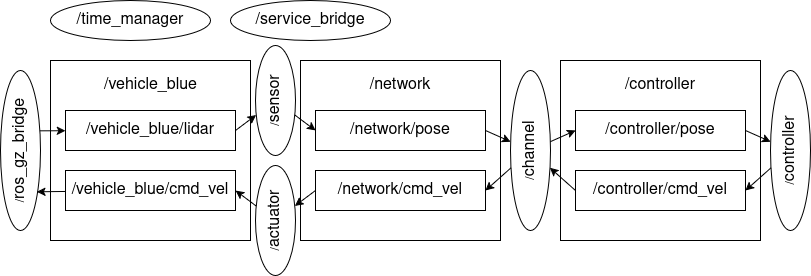}
    \caption{The interaction among the different nodes in the \gls{ros2} system.\vspace{-0.5cm}}
    \label{fig:ros2-nodes}
\end{figure}

In our simulations, we consider four nodes, as represented in Fig.~\ref{fig:ros2-nodes} and described below.
\begin{itemize}
    \item \texttt{sensor}: This node simulates the LiDAR of the \gls{agv}, which is used for localization and navigation. It interacts with the Gazebo simulation to get the odometry of the \gls{agv}, and publishes it to the \texttt{/network/pose} topic.
    \item \texttt{controller}: This node implements the control system of the \gls{agv}, as described in Sec.~\ref{sub:control}. 
    Specifically, it  subscribes to the \texttt{/controller/pose} topic to get the \gls{lidar} measurements of the \gls{agv}, and publishes the velocity commands that minimize the error between the desired and actual trajectories of the \gls{agv} to the \texttt{/controller/cmd\_vel} topic.
    \item \texttt{network}: This node simulates the communication between the different nodes in the system. It subscribes to the \texttt{/network/pose} topic to get the \gls{lidar} measurements of the \gls{agv}, and publishes them to the \texttt{/controller/pose} topic. In the other direction, it subscribes to the \texttt{/controller/cmd\_vel} topic to get the velocity commands for the \gls{agv}, and publishes them to the \texttt{/network/cmd\_vel} topic.
    The \texttt{network} node simulates the communication delay and \gls{prr}, based on the two-state Markov Process described in Sec.~\ref{sub:channel-model}. 
    \item \texttt{time\_manager}: This node is used to synchronize \mbox{\gls{ros2}} and Gazebo. It subscribes to the \texttt{/clock} topic to get the current simulation time, and controls the flow of the simulation by interacting with Gazebo accordingly.
    Specifically, the time manager keeps the list of the upcoming events in \gls{ros2}, and resumes the Gazebo simulation every time an event is solved. In other words, the time manager ensures that Gazebo does not overtake the \gls{ros2} simulation.
\end{itemize}

\section{Performance Evaluation}
\label{sec:perf-eval}
In Sec.~\ref{sub:params} we present our simulation parameters, and in Sec.~\ref{sub:results} we show our numerical results. 

\subsection{Simulation Parameters}
\label{sub:params}

\paragraph{AGV model}
We run simulations of $62$ s, where the \gls{agv} is forced to follow an eight-shape trajectory $p^*$ with amplitudes  $A=B=20$ m in both the $x$ and $y$ directions, moving across a ground plane of size $25\text{ m} \times25$ m.
The main parameters of the AGV model described in Sec.~\ref{sub:scenario} are $\rho = 1.2$ m, $O_l = (-0.5, 0.6)$, $O_r = (-0.5, -0.6)$, and $C = (-0.23, 0)$, where coordinates are relative to the center of the~chassis.

\paragraph{Controller}
The \gls{agv} receives periodic driving instructions from the control unit. The PID terms (P, I, D) have been found empirically via trial and error. Inspired by standard techniques for PID optimization, we first increased the proportional gain (P) until the system responded with noticeable oscillations. Then, we introduced the derivative term (D) to reduce the oscillatory behavior. Finally, the integral term (I) was increased to eliminate steady-state errors and ensure convergence to the desired trajectory.
The Gazebo engine operates at a fixed {update rate} $R=1000$ Hz, resulting in a frame duration of \SI{1}{ms}.

    
\paragraph{Metrics}
We evaluate via simulations the impact of the delay $\delta$ and PRR on the normalized control error function $\varepsilon$ expressed in Eq.~\eqref{eq:normalized_error}.



\subsection{Simulation Results}
\label{sub:results}

\label{sec:results}
\begin{figure}[b!]
    \centering
    \input{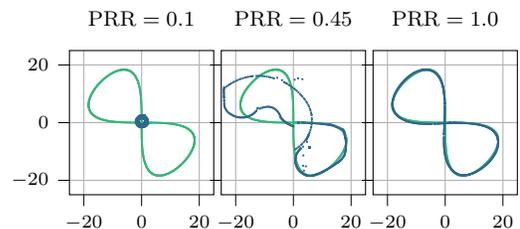}
    \caption{Trajectories (planned in green, actual in blue) of the AGV, vs. the PRR, for $\delta=0$~ms.}
    \label{fig:trajectories_packet_loss}
\end{figure}

In this initial set of results, we force the channel in one of the states ($G$ or $B$). 
First, in Fig.~\ref{fig:trajectories_packet_loss}, we set $\delta=0$ ms, and depict the optimal/planned path $p^*$ (in green) and the actual path $p$ (in blue) of the AGV in the $x$-$y$ plane as a function of the PRR, so as to visualize the performance of the control system.
We observe that the deviation of $p$ from $p^*$ increases as the PRR decreases, which demonstrates the non-negligible effect of the communication network on the control performance. 
For moderate packet loss ($\text{PRR}=0.45$), the \gls{agv} can generally follow the desired trajectory, although tracking deviations may occur during sharp turns, such as around $(x, y) = (-10,5)$. In this case, the AGV may hit the boundaries (walls) of the ground plane if control commands are lost, such as around $(x, y) = (-20, 15)$, which takes time to reorient to the original track.
When $\text{PRR}=0.1$, the \gls{agv} fails to follow $p^*$ entirely. Instead, it usually performs a series of circular movements, and is unable to reach the desired destination. This is due to the fact that the \gls{agv} receives few commands from the central unit, and continues to operate with constant linear and angular velocities. As a result, it tends to remain near the center of the world frame.

\begin{figure}[t!]
    \centering
    \input{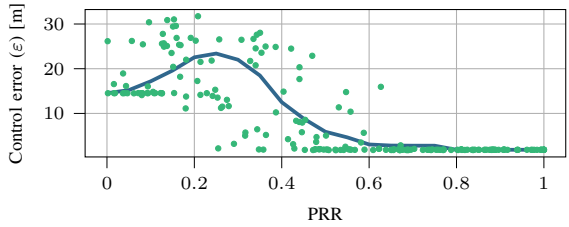}
    \caption{Average control error per simulation $\varepsilon$ (dots) and interpolation (line) vs. the \gls{prr}, for $\delta=0$~ms.}
    \label{fig:PRR_MSE}
\end{figure}

\begin{figure}[t!]
    \centering
    \input{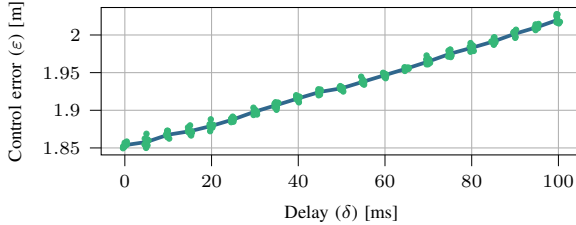}
    \caption{Average control error per simulation $\varepsilon$ (dots) and interpolation (line) vs. $\delta$, for $\text{PRR}=1$.\vspace{-0.5cm}}
    \label{fig:Delay_MSE}
\end{figure}

To formally evaluate the impact of the communication parameters on the control error, in Figs.~\ref{fig:PRR_MSE} and~\ref{fig:Delay_MSE} we plot $\varepsilon$ as a function of the PRR and $\delta$, respectively.
As expected, the average value of $\varepsilon$ decreases as the PRR ($\delta$) increases (decreases), which is in line with the qualitative trends observed in Fig.~\ref{fig:trajectories_packet_loss}.
In particular, the dependence of $\varepsilon$ on the PRR in Fig.~\ref{fig:PRR_MSE} has a bell shape. When $\text{PRR} < 0.15$, the AGV is unable to follow the desired trajectory, and remains near the center of the ground plane. In this condition, the AGV is approximately equidistant from most points in the environment, and $\varepsilon~\simeq 15$ m, on average.
For $0.15 < \text{PRR} < 0.45$, the average error is up to $\varepsilon\simeq23$ m, i.e., when the AGV is at the boundaries of the ground plane, at distant locations from the target path.
Finally, when $\text{PRR}>0.45$, the performance of the control system improves, even though the average $\varepsilon\neq 0$ m, as it appears from the rightmost part of Fig.~\ref{fig:PRR_MSE}.
This is due to the simple design of the control system, and to the noise introduced by the Gazebo simulation.

As far as $\delta$ is concerned, in Fig.~\ref{fig:Delay_MSE} we show that $\varepsilon$ grows almost linearly with the average delay. 
Still, the range of $\varepsilon$ is limited between $1.8$ and $2.1$ m, in contrast to the range for the PRR in Fig.~\ref{fig:PRR_MSE} where $\varepsilon$ grows up to $30$ m. This result suggests that packet loss is the dominant component for the control error.
Indeed, while $\delta$ may cause the AGV to oscillate if control commands are delayed, it does not generally result in a complete loss of~trajectory.

\begin{figure}[t!]
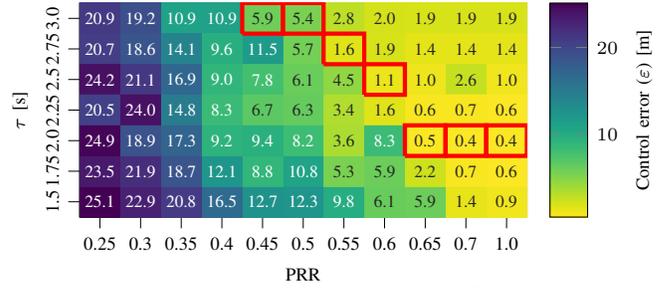

    \centering
    \include{tikz/heatmap}
    \vspace{-0.7cm}
    \caption{Average control error $\varepsilon$ as a function of the \gls{prr} and the look-ahead parameter $\tau$. Red boxes indicate the (\gls{prr}, $\tau$) pairs that minimize $\varepsilon$. 
    \vspace{-0.5cm} }
    \label{fig:heatmap_prr_tau}
\end{figure}

Communication and control performance is affected by other parameters, especially $\tau$.
As described in Sec.~\ref{sub:control}, the controller computes the driving instructions to minimize the angular and linear errors between $p$ and $p^*$ by ``looking-ahead'' along $p^*$ by $\tau$ time units: increasing $\tau$ allows the AGV to anticipate upcoming changes in the path, while decreasing $\tau$ provides  more accurate tracking, though only relative to the current segment of the path.
Fig.~\ref{fig:heatmap_prr_tau} shows in a heatmap the average $\varepsilon$ for different values of \gls{prr} and $\tau$.
We observe that $\tau$ has a significant impact on $\varepsilon$. For each column, we mark with a red box the pair (\gls{prr}, $\tau$) that minimizes $\varepsilon$. We can see that the optimal value of $\tau$ is not constant, but depends on the \gls{prr}. 
Notably,  $\tau$ should be increased when the PRR decreases; for example, $\tau=3$ s when $\text{PRR}=0.45$, vs. $\tau=2$ s when $\text{PRR}=0.65$. 
This is due to the fact that the controller can predict further along $p^*$, generating commands that remain effective even if future updates are lost or delayed. Under these conditions, precise short-term tracking is unreliable, and the system tends to 
adopt a more predictive control strategy.
Interestingly, for $\text{PRR}=1$, $\varepsilon$ initially decreases as $\tau$ decreases, but starts increasing again when $\tau<2$ s. This is likely because, with very small values of $\tau$, driving instructions may change too frequently for the AGV to react effectively, especially at high speeds.
These results demonstrate the ability of the controller to compensate for the packet loss by adjusting $\tau$, which validates the accuracy of our control framework. Still, the overall $\varepsilon$ increases as the PRR decreases, which is consistent with our previous observations in Fig.~\ref{fig:PRR_MSE}.

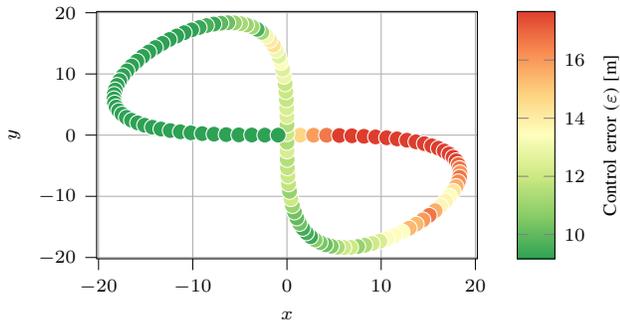
\begin{figure}[t!]
    \centering
\begin{tikzpicture}

  \definecolor{darkgray176}{RGB}{176,176,176}
  \definecolor{darkkhaki183224117}{RGB}{183,224,117}
  \definecolor{firebrick165038}{RGB}{165,0,38}
  \definecolor{forestgreen010455}{RGB}{0,104,55}
  \definecolor{tomato24210665}{RGB}{242,106,65}
  
  \definecolor{color0}{RGB}{45,161,84}
  \definecolor{color1}{RGB}{134,203,102}
  \definecolor{color2}{RGB}{205,233,131}
  \definecolor{color3}{RGB}{254,254,189}
  \definecolor{color4}{RGB}{253,210,127}
  \definecolor{color5}{RGB}{248,139,81}
  \definecolor{color6}{RGB}{221,61,45}
  
  \begin{axis}[
  width=0.75\columnwidth,
height=0.55\columnwidth,
  legend style={fill opacity=0.8, draw opacity=1, text opacity=1, draw=none},
  tick align=outside,
  tick pos=left,
  x grid style={darkgray176},
  xlabel={$x$},
  xmajorgrids,
  xmin=-20.204485737382, xmax=20.2079581338087,
  xminorgrids,
  xtick style={color=black},
  y grid style={darkgray176},
  ylabel={$y$},
  ymajorgrids,
  ymin=-20.2075685715192, ymax=20.2058428576499,
  yminorgrids,
  ytick style={color=black},colorbar,
  colormap={hot}{
      color=(color0) color=(color1) color=(color2) color=(color3) color=(color4) color=(color5) color=(color6)
  },
  colorbar style={
      ylabel={Control error ($\varepsilon$) [m]}
      }, 
  point meta min=9.162776349400474, point meta max=17.67189339786439, 
  ]
  \addplot[
    scatter,
    only marks,
    scatter src=explicit,
    mark=*,
    mark size=3,
    scatter/use mapped color={
      draw=white,
      fill=mapped color,
    }]
  table[meta=meta]{
  x  y  meta
  0.00 0.00 13.15
  1.41 -0.00 14.72
  2.82 -0.01 15.90
  4.20 -0.02 16.61
  5.56 -0.06 18.56
  6.89 -0.11 20.17
  8.17 -0.19 19.97
  9.40 -0.29 21.22
  10.58 -0.43 21.99
  11.69 -0.61 21.74
  12.73 -0.83 20.35
  13.69 -1.09 21.28
  14.58 -1.39 21.32
  15.37 -1.75 22.67
  16.08 -2.14 21.87
  16.69 -2.59 20.54
  17.21 -3.08 18.35
  17.64 -3.61 18.22
  17.96 -4.19 18.20
  18.19 -4.81 16.92
  18.33 -5.47 16.71
  18.37 -6.16 16.13
  18.32 -6.89 16.66
  18.18 -7.64 16.31
  17.96 -8.40 15.54
  17.65 -9.19 14.16
  17.27 -9.98 13.33
  16.82 -10.78 13.73
  16.31 -11.57 13.67
  15.73 -12.35 15.56
  15.10 -13.11 16.64
  14.43 -13.85 15.97
  13.72 -14.55 15.34
  12.98 -15.21 15.36
  12.22 -15.83 13.41
  11.44 -16.40 13.45
  10.64 -16.90 13.36
  9.85 -17.34 12.28
  9.06 -17.71 12.09
  8.27 -18.00 11.34
  7.51 -18.21 10.89
  6.76 -18.34 11.99
  6.04 -18.37 12.04
  5.36 -18.31 11.33
  4.71 -18.16 10.55
  4.09 -17.92 10.17
  3.52 -17.57 10.40
  2.99 -17.13 10.66
  2.51 -16.60 9.44
  2.07 -15.97 9.57
  1.68 -15.24 9.52
  1.34 -14.43 10.26
  1.04 -13.54 10.01
  0.79 -12.56 10.28
  0.58 -11.51 10.99
  0.41 -10.39 12.01
  0.27 -9.20 10.91
  0.17 -7.96 11.76
  0.10 -6.67 11.99
  0.05 -5.34 11.77
  0.02 -3.97 11.53
  0.01 -2.58 12.35
  0.00 -1.18 11.37
  -0.00 0.24 11.93
  -0.00 1.65 11.45
  -0.01 3.05 12.37
  -0.03 4.43 12.30
  -0.06 5.79 11.86
  -0.12 7.11 11.94
  -0.20 8.38 11.91
  -0.32 9.61 12.89
  -0.46 10.77 12.45
  -0.65 11.87 13.48
  -0.87 12.90 12.53
  -1.14 13.85 13.04
  -1.45 14.72 14.36
  -1.81 15.50 14.09
  -2.21 16.19 13.06
  -2.67 16.79 11.14
  -3.16 17.29 9.67
  -3.71 17.70 11.11
  -4.29 18.01 11.03
  -4.92 18.22 11.26
  -5.58 18.34 10.90
  -6.28 18.37 10.19
  -7.01 18.30 9.95
  -7.76 18.15 9.37
  -8.54 17.91 9.62
  -9.32 17.59 9.37
  -10.12 17.20 8.91
  -10.91 16.74 7.21
  -11.70 16.21 6.56
  -12.48 15.63 5.88
  -13.23 14.99 6.09
  -13.97 14.32 5.99
  -14.66 13.60 5.75
  -15.32 12.86 6.52
  -15.93 12.09 5.92
  -16.49 11.30 5.68
  -16.98 10.51 6.16
  -17.41 9.71 5.73
  -17.76 8.92 5.85
  -18.04 8.14 6.23
  -18.24 7.38 5.89
  -18.35 6.64 6.30
  -18.37 5.93 6.94
  -18.30 5.25 6.98
  -18.13 4.60 6.71
  -17.86 3.99 5.47
  -17.51 3.43 5.46
  -17.05 2.91 4.74
  -16.50 2.43 4.16
  -15.85 2.00 5.27
  -15.11 1.62 5.61
  -14.29 1.29 5.32
  -13.38 1.00 5.75
  -12.39 0.75 4.82
  -11.32 0.55 5.73
  -10.19 0.38 5.90
  -9.00 0.25 6.08
  -7.75 0.16 5.25
  -6.45 0.09 4.29
  -5.11 0.04 7.10
  -3.74 0.02 6.93
  -2.35 0.00 6.85
  -0.94 0.00 5.68
  };
  \end{axis}
  
  \end{tikzpicture}
  
    \vspace{-0.7cm}
    \caption{Spatial distribution of the control error $\varepsilon$ in the x-y plane when the channel is in the ``bad'' ($B$) state. Regions with more intense colors indicate higher control errors. 
    \vspace{-0.6cm}}
    \label{fig:risk_map}
\end{figure}

Finally, we consider the two-state Markov Process channel described in~Sec.~\ref{sub:channel-model}. Specifically, the stationary probabilities are set to $P[G]=0.75$ and $P[B]=0.25$, so that the network is in the $G$ state most of the time.  In the $G$ state, we set $\delta=0$ ms and $\text{PRR}= 1$. In the $B$ state, we set $\delta=500$ ms and $\text{PRR}=0.5$.
In Fig.~\ref{fig:risk_map} we plot the position of the AGV in the $x$-$y$ plane, where the color of each point indicates the average value of $\varepsilon$ when the network is in the $B$ state at that point in the space. 
Specifically, the \gls{agv} starts at $(x,y)=(0,0)$, and moves along an eight shape heading toward positive values of $x$. 
Points in the path are sampled at uniform time intervals, so the density of points is inversely proportional to the speed of the AGV.
The results show that, in some path segments, $\varepsilon$ is large and the \gls{agv} deviates significantly from the desired trajectory, especially before it turns or slows down, due to the negative effect of the channel in the $B$ state.
From a qualitative point of view, during these critical movements, the AGV relies on frequent control commands to adjust its speed and direction. Under poor network conditions, if these commands are delayed or lost, the AGV may take an unintended trajectory that the control system cannot fully mitigate or correct.
This hypothesis is confirmed by the fact that $\varepsilon$ decreases after the first and second turns, at around $(x,y)=(17,-10)$ and $(12,-15)$, when the AGV accelerates in a straight line. On the contrary, it increases before the first, second and third turns, from $(0,0)$ to $(19,-3)$, from $(18,-8)$ to $(12,-15)$, and from $(0,5)$ to $(-2,15)$, respectively, when the AGV slows~down.

\section{Conclusions}
\label{sec:conclusions}
In this paper we presented a simulation framework to evaluate the relationship between the communication and the control system of an industrial \gls{ncs} responsible for guiding an \gls{agv} along a predefined trajectory. The framework consists of a Gazebo module that simulates the physics of the \gls{agv}, and a \gls{ros2} module that implements the control system.
We measured the impact of the network performance, in terms of communication delay and PRR, on the performance of the control system, measured by the error between the planned and actual path of the AGV.
We show that the PRR has a stronger influence on the control error than the delay, even though it is possible to mitigate this effect by acting on the controller's parameters, such as changing $\tau$ for different values of \gls{prr}.
Finally, we showed that there are some points in the path where the \gls{agv} is prone to instability, especially before it turns or slows down to drive into a curve. This means that the control system is more sensitive to network errors in those regions, which motivates more advanced control strategies. 

However, we acknowledge some limitations in our approach. First, the two-state Markov channel model, while useful to capture stochastic network behaviors, does not account for more complex real-world phenomena such as bursty interference or spatial correlation in packet loss. Future work will consider a more refined model to enhance realism. Moreover, a PID-based control strategy may be suboptimal in the presence of highly dynamic network conditions. Advanced control strategies such as Reinforcement Learning (RL) may be required to further increase robustness and adaptability.
Finally, as part of our future work, we plan to implement a practical demonstrator using commercial AGVs operating on an industrial floor, to validate the simulation results of this paper under real-world conditions.

\section*{Acknowledgement}
This work was carried out in the framework of the CNIT National Laboratory WiLab and the WiLab-Huawei Joint Innovation Center. This work was also partially supported by the European Union under the Italian National Recovery and Resilience Plan (NRRP) Mission 4, Component 2, Investment 1.3, CUP C93C22005250001, partnership on ``Telecommunications of the Future'' (PE00000001 - program ``RESTART'').

\bibliographystyle{IEEEtran}
\bibliography{bibliography.bib}

\end{document}